# A decentralized future for the open-science databases

Gaurav Sharma
Department of Biotechnology, Indian Institute of Technology Hyderabad, Sangareddy, Telangana, India 502284
sharmaG@bt.iith.ac.in
ORCiD: 0000-0002-2861-7446

Viorel Munteanu
Department of Computers, Informatics and Microelectronics, Technical University of Moldova, Chisinau, 2045, Moldova
Department of Biological and Morphofunctional Sciences, College of Medicine and Biological Sciences, University of Suceava, Suceava, 720229, Romania
viorel.munteanu@lt.utm.md
ORCiD: 0000-0002-4133-5945

Nika Mansouri Ghiasi
Department of Information Technology and Electrical Engineering, ETH Zurich, Switzerland
n.mansorighiasi@gmail.com
ORCiD: 0000-0002-0833-0042

Jineta Banerjee
Sage Bionetworks, Washington, USA
jineta.banerjee@sagebionetworks.org
ORCiD: 0000-0002-1775-3645

Susheel Varma
Sage Bionetworks, Washington, USA
susheel.varma@sagebase.org
ORCiD: 0000-0003-1687-2754

Luca Foschini
Sage Bionetworks, Washington, USA
luca.foschini@sagebase.org

Kyle Ellrott
Oregon Health and Science University, Portland, Oregon
ellrott@ohsu.edu

Onur Mutlu
Department of Information Technology and Electrical Engineering, ETH Zurich, Switzerland
omutlu@gmail.com
ORCiD: 0000-0002-0075-2312

Dumitru Ciorbă
Department of Computers, Informatics and Microelectronics, Technical University of Moldova,

Chisinau, 2045, Moldova
dumitru.ciorba@fcim.utm.md
ORCiD: 0000-0002-3157-5072

Roel A. Ophoff
Center for Neurobehavioral Genetics, Semel Institute for Neuroscience and Human Behavior, University of California Los Angeles, Los Angeles, CA 90095, USA.
ophoff@g.ucla.edu
ORCiD: 0000-0002-8287-6457

Viorel Bostan
Department of Computers, Informatics and Microelectronics, Technical University of Moldova, Chisinau, 2045, Moldova
Email: viorel.bostan@adm.utm.md
ORCiD: 0000-0002-2422-3538

Christopher E Mason
Department of Physiology and Biophysics, Weill Cornell Medicine, New York, NY, 10065, USA
WorldQuant Initiative for Quantitative Prediction, Weill Cornell Medicine, New York, NY, 10065, USA,
Email: chm2042@med.cornell.edu
ORCID: https://orcid.org/0000-0002-1850-1642

Jason H. Moore
Department of Computational Biomedicine, Cedars-Sinai Medical Center, Los Angeles, CA 90069
Jason.Moore@csmc.edu
ORCiD: 0000-0002-5015-1099

Despoina Sousoni
ELIXIR Hub, Wellcome Genome Campus, Hinxton, Cambridge, United Kingdom
despoina.sousoni@elixir-europe.org
ORCiD: 0000-0002-5426-7881

Arunkumar Krishnan
Department of Biological Sciences, IISER Berhampur, Odisha, India
akrishnan@iiserbpr.ac.in
ORCiD: 0000-0002-9677-9092

Christopher E. Mason
Department of Physiology and Biophysics, Weill Cornell Medicine, New York, NY, USA
The HRH Prince Alwaleed Bin Talal Bin Abdulaziz Alsaud Institute for Computational Biomedicine, Weill Cornell Medicine, New York, NY, USA
The WorldQuant Initiative for Quantitative Prediction, Weill Cornell Medicine, New York, NY, USA

The Feil Family Brain and Mind Research Institute, Weill Cornell Medicine, New York, NY, USA
chm2042@med.cornell.edu
ORCiD: 0000-0002-1850-1642

Mihai Dimian
Department of Computers, Electronics and Automation, Stefan cel Mare University of Suceava, 720229 Suceava, Romania
dimian@usm.ro
ORCiD: 0000-0002-2093-8659

Gustavo Stolovitzky
Department of Pathology, NYU Grossman School of Medicine, New York, USA
Gustavo.Stolovitzky@nyulangone.org
ORCiD: 0000-0002-9618-2819

Fabio G. Liberante
ELIXIR Hub, Wellcome Genome Campus, Hinxton, Cambridge, United Kingdom
fabio.liberante@elixir-europe.org
ORCiD: 0000-0002-0192-5385

Taras K. Oleksyk
Department of Biological Sciences, Oakland University, Rochester, MI 48309 USA
Department of Biology, and Department of Biological Sciences, Uzhhorod National University, Uzhhorod 88000, Ukraine
oleksyk@oakland.edu
ORCiD: 0000-0002-8148-3918

Serghei Mangul
Sage Bionetworks, Washington, USA
serghei.mangul@gmail.com
ORCiD: 0000-0003-4770-3443

Correspondence to sharmaG@bt.iith.ac.in and serghei.mangul@gmail.com

**Abstract:**

The continuous and reliable open access to curated biological data repositories is indispensable for accelerating rigorous scientific inquiry and fostering reproducible research outcomes. However, the current paradigm, which relies heavily on centralized infrastructure for the storage and distribution of foundational biomedical datasets, inherently introduces significant vulnerabilities. This centralized model is susceptible to single points of failure, including cyberattacks, technical malfunctions, natural disasters, and even political or funding uncertainties. Such disruptions can lead to widespread data unavailability, data loss, integrity compromises, and substantial delays in critical research, ultimately impeding scientific progress. The downstream effect of such interruptions can be the widespread paralysis of diverse research activities, including computational, clinical, molecular, and climate studies. This scenario vividly illustrates the inherent dangers of consolidating essential scientific resources within a single geopolitical or institutional locus. As data generation is accelerating and the global landscape continues to fluctuate, the sustainability of centralized models must be critically re-evaluated. A shift toward federated and decentralized architectures may offer a robust and forward-looking approach to enhancing the resilience of scientific data infrastructures by reducing exposure to governance instability, infrastructural fragility, and funding volatility, while also promoting equity and global accessibility. Inspired by established models such as ELIXIR's federated infrastructure and the policy and funding frameworks developed by CODATA and the Global Biodata Coalition (GBC), emerging Decentralized Science (DeSci) initiatives can contribute to building more resilient, fair, and incentive-aligned data ecosystems. The future of open science depends on integrating these complementary approaches to establish a globally distributed, economically sustainable, and institutionally robust infrastructure that safeguards scientific data as a public good, further ensuring continued accessibility, interoperability, and preservation for generations to come. Here, we examine the structural limitations of centralized repositories, evaluate federated and decentralized models, and propose a hybrid framework for resilient, fair, and sustainable scientific data stewardship.

**Introduction:**

Scientific research has traditionally relied on funding from national governments, making it inherently susceptible to changes in governmental priorities and resource allocation. Decisions on resource allocation are shaped by policymakers' perceptions of research value, their commitment to evidence-based decision-making, and broader ideological frameworks that define national scientific agendas. However, because scientific research is fundamentally rooted in critical analysis and empirical evidence, it sometimes comes into conflict with political ideologies, particularly when findings challenge dominant narratives or economic interests.[1] This tension is often reinforced by a misconception regarding the operational nature of science, with policymakers expecting immediate, predictable, and definitive outcomes rather than the iterative and uncertain nature of discovery. The mismatch between short-term political cycles and the long timescales required for scientific progress leads to funding volatility, often resulting in missed opportunities for transformative breakthroughs. Fluctuations in governmental support, changes in political leadership, and evolving legislative priorities may significantly impact the direction, scope, and sustainability of scientific investigations. These vulnerabilities are particularly pronounced for scientific data infrastructure that relies on centralized repositories - architectures in which storage, access, and governance are managed by a single institution or authority. While strong governmental commitment can accelerate discovery, funding cuts and policy shifts may disrupt critical research,[2–5] delay technological advancements, and compromise long-term scientific progress.[6,7]

Crucially, the continued advancement of scientific progress has been significantly enabled by the growing availability of federally funded open-access data. which over the past several decades has enabled researchers worldwide to mine, analyze, and repurpose these vast public datasets to advance understanding of life sciences, and its applications in human health.[8–10] Open-access data now plays a central role across the biomedical sciences, and its unrestricted availability has accelerated scientific discovery, fostered global collaboration, and enabled researchers across disciplines to build upon existing knowledge without geographic barriers.[11] An example of the socio-economic value of freely accessible databases in the life sciences is provided by the study on the cost-benefit analysis of UniProt, an essential protein database for researchers globally.[12] This study shows that UniProt's total annual costs amount to approximately €14.6 million, and the database generates benefits ranging from €373 million to €565 million annually. These calculations do not capture UniProt's pivotal role in advancing new AI technologies such as AlphaFold, which have become indispensable in the daily practices of life scientists. While “open-access” traditionally referred to fully public datasets with unrestricted availability, the definition has evolved in response to the sensitive nature of certain types of health data. The current understanding of open access includes resources such as the UKBiobank, which provides tiered access, offering some data openly while restricting more sensitive modalities to approved users under specific access conditions academic and commercial users under specific access conditions for health-related research.[13,14] Even with these constraints, such datasets are widely recognized as part of the open science ecosystem.

Scientific advances enabled by open-access data now face growing risk, as emerging constraints on data sharing, driven by changes in governance structures, institutional policies, and increasing infrastructure centralization, threaten to restrict access to public repositories, disrupt knowledge exchange, widen disparities in research capacity, and concentrate control within a small number of well-resourced institutions and nations.[15–18] Moreover, the growing risk of monetizing access, with databases potentially placed behind paywalls and governments or institutions imposing access fees on specific countries or groups, may hinder the free exchange of scientific knowledge, disproportionately disadvantaging researchers in resource-limited settings and deepening global inequalities in scientific progress.[19–21] These trends threaten to erode the foundational principles of open science by restricting participation, limiting reproducibility, and reinforcing structural asymmetries in who can generate, access, and apply scientific data. Addressing these challenges requires a sustained commitment to preserving open-access scientific data as a global public good by ensuring that knowledge databases remain publicly accessible, protected from tampering, and resistant to institutional or political interference. This, in turn, safeguards the integrity and continuity of research while promoting a fairer and collaborative scientific ecosystem that drives innovation and advances discoveries for the collective benefit of society.[15,22]

A prominent example of biological global data-sharing infrastructure is the International Nucleotide Sequence Database Collaboration (INSDC),[23] a long-standing alliance among the National Library of Medicine's National Center for Biotechnology Information (NLM-NCBI),[24] the European Bioinformatics Institute at the European Molecular Biology Laboratory (EMBL-EBI),[25] and the National Institute of Genetics under the Research Organization of Information and Systems (ROIS-NIG) in Japan. These institutions jointly maintain synchronized repositories, including GenBank and the Sequence Read Archive (SRA) at NCBI,[25,26] the European Nucleotide Archive (ENA) at EMBL-EBI,[27] and DDBJ at ROIS-NIG,[28] which exchange nucleotide sequence data on a daily basis, thereby promoting broad, open access to foundational biological information. However, despite these coordinated efforts, only a fraction of genomic, proteomic, and structural biology data is fully synchronized across these platforms. Due to the massive scale and complexity of synchronization, key repositories such as SRA, ClinVar, and more than 20 other NCBI-managed resources, alongside literature databases, including PubMed and Europe PMC, remain only partially shared, leading to gaps in data availability, risks of fragmentation, and persistent inequities in access, highlighting the limitations of past data-sharing initiatives and the challenges of maintaining real-time, federated updates. Recent reports documenting the removal or alteration of scientific content from federal government web pages in response to political pressures [15] highlight the urgent need for a resilient and interoperable data infrastructure that ensures efficient integration, fair access, and long-term stability of publicly funded scientific resources. As scientific data expands in scale and complexity, addressing these challenges is becoming increasingly critical to sustaining global research progress and maximizing the impact of open-access resources.

**Limitations of Centralized Data Repositories:**

Centralized data repositories provide a structured and convenient framework for storing and managing large-scale biological datasets by facilitating standardization, ensuring version control, and supporting efficient data access. However, this architectural model can obscure systemic risks that threaten the sustainability of open science. As data volume, complexity, and generation rates increase, these systems face growing challenges, including latency, reduced efficiency, and infrastructure bottlenecks, while remaining fundamentally constrained by scalability, performance, and accessibility limitations.[29,30] This inherent fragility often leads to a "cycle of doom", where efforts to improve the performance through added layers of complexity introduce new inefficiencies, further exacerbating system-wide limitations. The reliance on a single, centralized infrastructure also constrains computational resources, forcing all users to compete within the same system and amplifying bottlenecks in high-demand scenarios. [31]

This growing structural and functional rigidity renders centralized repositories increasingly susceptible to single points of failure, where technical malfunctions, cyberattacks, or institutional/regional/national disruptions may compromise access to entire datasets.[32,33] Concentrating critical data within a single infrastructure heightens exposure to security breaches[33–36] and the potential misuses of sensitive information.[37–39] The high cost of establishing and maintaining centralized repositories can create barriers to entry for smaller institutions or those from under-resourced communities, leading to a loss of diverse voices and perspectives in scholarly discourse and reinforcing existing inequalities.[40] Additionally, restrictive data access policies, licensing fees, or geographic limitations may disproportionately affect researchers from low-resource settings, limiting their ability to contribute to and benefit from shared data.[41–43] Another challenge can be that centralized systems may not easily accommodate novel data structures or emerging analytical approaches, potentially stifling innovation.[44,45]

Furthermore, reliance on a single governing entity introduces systemic risks, as policy changes, funding cuts, or institutional biases may directly impact data availability and governance.[46,48,49] Data governance in centralized repositories is often dictated by a small group of stakeholders, leading to ethical concerns regarding data ownership, control, and decision-making.[47] Researchers contributing data to these repositories may have limited say in how their data is used, who has access, or whether data can be withdrawn. These challenges are particularly pronounced in cases involving sensitive or personally identifiable data, where the lack of robust governance frameworks can lead to ethical dilemmas and non-compliance with data protection regulations. If funding is reduced or redirected, repositories may face operational difficulties, leading to data loss, restricted access, or the need for migration to alternative platforms. This dependency creates long-term sustainability concerns, particularly for repositories that lack clear financial models for continued maintenance and growth.

A lack of governmental mandate for organized data collection and deposition to a single repository for sharing has given rise to multiplicity and fragmentation **(Figure 1)**. The proliferation of data repositories has become commonplace, with each repository uniquely designed to accommodate specific data sharing goals and specialized capabilities for storing diverse data modalities.[50–53] This multitude of repositories grew organically at various points in time in response to the need for organized data collection to answer specific questions. While a strategy to decentralize data sharing initially seemed promising, it inadvertently created significant hurdles in data sharing and reuse due to a lack of interoperability between multiple repositories, underscoring the critical need for a more thoughtful approach to decentralization[54]

**Strategies for decentralized scientific data ecosystems:**

Open and interoperable infrastructures are essential for the continuity of scientific progress and effective use of shared data, but their maintenance imposes substantial and ongoing costs, and the question of long-term financial responsibility remains largely unanswered. To introduce the major concepts discussed in this paper, we define three distinct types of architecture in which scientific data are stored, accessed, and governed. These concepts are visually illustrated in Figure 1. Centralized architectures (Figure 1A) locate the entire repository and decision-making authority within a single institution, yielding uniform standards but creating a single point of technical or administrative failure. By contrast, Federated architectures (Figure 1B) maintain multiple, sovereign data nodes that interoperate through shared metadata, policy, and security standards, distributing risk while retaining cohesion. Finally, Decentralized Science or "DeSci" architectures (Figure 1C) disaggregate storage, provenance, and incentive mechanisms across many independent nodes that are linked by ledgers and distributed file systems, aiming for maximal resilience, transparency, and community engagement. Understanding these three architectural concepts is crucial, as they offer varying levels of control, scalability, and fairness, thereby aiding in determining the most suitable model for diverse data needs. Federated models, such as those implemented across Europe, including the EOSC,[55] ELIXIR,[56] and BBMRI-ERIC,[57] offer structural resilience by enabling continuity even when individual national contributions fluctuate. Below, we highlight a few initiatives that reflect diverse models of distributed governance and coordination, ranging from structured international collaborations like ELIXIR and CODATA to emerging DeSci efforts.

**1. Lessons from a federated network like ELIXIR:**

One key example of a significant shift from traditional centralized data repositories toward distributed peer-managed systems is the European life-sciences infrastructure for biological information, also known as ELIXIR, established over a decade ago as a network of independently operated resources and services provided by its 25 nodes (22 members and EMBL, a non-country node; and additional 3 observers). Rather than functioning as a strict singular federated infrastructure, ELIXIR represents a collaboration among diverse entities working collectively to maintain open, secure, and FAIR-compliant (Findable, Accessible, Interoperable, and Reusable) life science data and bioinformatics resources.[58] This pan-European network, coordinated through a central hub at the Wellcome Genome Campus in Hinxton, UK, operates by aligning independent resources and capabilities spread across 25 European countries (including Observers) and approximately 240 research institutes. Each node, whether a national or institutional entity, maintains autonomy and manages its own specialized databases and resources across fields such as genomics, proteomics, metabolomics, and computational biology.[56] This distributed arrangement supports long-term data preservation, ensures reliable, uninterrupted data access, and promotes inclusivity in the scientific community by mitigating risks associated with centralized control or censorship.

ELIXIR operates under a federated data architecture, eschewing centralized data warehousing in favor of a distributed network in which each national or institutional node retains sovereign control over its datasets. This autonomy is balanced by tight coordination through ELIXIR's central registry, unified APIs, and metadata indexing services, enabling efficient data discovery, retrieval, and integration. Critically, the ELIXIR model relies on uniform standards for data and metadata, developed through community consensus, ensuring interoperability and supporting data exchange across biological domains.[58] An exception is the set of ELIXIR Deposition Databases, mostly operated by the EMBL-EBI, which itself is an intergovernmental organization that is funded by a number of member states and therefore not beholden to a particular nation.

Through its established standards, platforms, and partnerships, ELIXIR coordinates national efforts into a unified European response, integrating data repositories, disseminating analysis workflows via platforms such as Galaxy (https://elixir-europe.org/about/groups/galaxy-wg), and harmonizing metadata to ensure cross-domain usability. This standardized approach has proven highly effective for large-scale, pan-European projects, such as the European '1+ Million Genomes' Initiative, which aims to enable secure access to high-quality genomic and clinical data.[59] Additionally, the agility of this approach was demonstrated during the COVID-19 pandemic when ELIXIR rapidly mobilized distributed resources and expertise through the COVID-19 Data Portal to enable rapid sharing and analysis of high-quality, open-source, and shareable biomedical datasets. This pan-European resource, supported by EMBL-EBI and ELIXIR partners, exemplifies the potential of a distributed infrastructure to support large-scale collaborative research during potential crises.[60]

The success of ELIXIR's distributed infrastructure offers a model that can inform the development of other global data infrastructures (also see the case study of Federated EGA (FEGA) in Box 1, which collaborates with ELIXIR extensively). For example, using multi-node data hosting combined with strategic public-private partnerships with technology providers and research institutions, core biological databases could be mirrored and maintained across multiple geographical regions. This would significantly improve system resilience, reduce reliance on a single point of control, and safeguard data continuity in the face of political, financial, or infrastructural disruptions, ensuring continuous access to vital resources. Establishing such international partnerships analogous to ELIXIR's country-based nodes would foster a more sustainable and resilient approach to data management, strengthening global open science collaboration and mitigating the risks associated with centralized governance.

**Box 1. Case Study: from European Genome-phenome Archive (EGA) to Federated EGA (FEGA)**:

The European Genome-phenome Archive (EGA), initially launched in 2008 by the European Bioinformatics Institute (EMBL-EBI) and subsequently co-managed with the Centre for Genomic Regulation (CRG) since 2012, began as a centralized repository for securely storing and distributing sensitive human genomic and phenotypic data. As genomic data generation increased across Europe, centralized data management faced limitations due to varying national data privacy laws, particularly as the importance of individual privacy, national sovereignty and personalized medicine initiatives has increased in recent years. To address these challenges, the Federated EGA (FEGA) was established in 2022, transitioning from a centralized model to a fully federated network. In this model, each participating country operates its own node, storing and managing data locally while enabling metadata sharing across the network to facilitate international research collaboration. This federated approach, initiated with national nodes in Finland, Germany, Norway, Spain, and Sweden, has since expanded, significantly enhancing data accessibility, compliance with local regulations, and overall resilience. By allowing secure and compliant international data sharing, FEGA supports major European genomic research initiatives, exemplified by the '1+ Million Genomes' project.

## 2. CODATA as the infrastructure spine for cross-domain data futures:

CODATA (Committee on Data of the International Science Council) is an international organization established in 1966 to improve the quality, reliability, management, and accessibility of scientific data, whose core mission is to promote open, interoperable data management along with encouraging international collaboration. In addition to FAIR, CODATA also follows CARE (Collective Benefit, Authority to Control, Responsibility, and Ethics) principles in its integration of traditional knowledge, health data, and community-generated scientific contributions into global research infrastructures. CODATA supports diverse strategic initiatives, including long-term programs such as the Global Open Science Cloud (GOSC), which aims to establish accessible data-sharing frameworks across regional and national open science infrastructures. For example, GOSC facilitates interoperability between Europe's EOSC, China's CSTCloud, and Africa's initiatives. Simultaneously, CODATA works closely with international organizations such as the Research Data Alliance (RDA), World Data System (WDS), and UNESCO to advance globally coordinated principles of open access, fair data sharing, and transparent governance. To further strengthen global data interoperability, CODATA promotes the development of national nodes or data centres that comply with international standards, ensuring machine-readable metadata and enabling cross-border and cross-disciplinary data reuse.

### 3. Global Biodata Coalition (GBC) is coordinating funding for core biodata infrastructure:

A crucial but often underemphasized dimension of sustaining open-science data ecosystems is long-term financial viability. GBC directly addresses this challenge by bringing together the world's major research funders to coordinate support for critical biomedical data resources.[61] Recognizing that many foundational biomedical databases operate under chronic funding uncertainty, the GBC serves as a collaborative platform to identify and sustain a portfolio of Global Core Biodata Resources (GCBRs),[62] ensuring their continued integrity within the global research infrastructure. These resources are essential to both public and private research worldwide, yet many remain vulnerable to funding fluctuations, posing a systemic risk to data continuity and access. By establishing shared principles between database governance and funding agencies, assessment criteria, and funding models, the GBC promotes international alignment around the stewardship of biomedical data infrastructure. Although not decentralized in architecture, the GBC's globally distributed, multi-funder model introduces an essential governance and economic pillar that can complement technical decentralization. In this way, it offers a long-lasting pathway for securing the continuity of open-access scientific databases in an increasingly data-driven research landscape.

### 4. Embedding Decentralization into the fabric of scientific collaboration:

The emerging DeSci movement calls for a radical rethinking of scientific data curation, preservation, and sharing through globally distributed backups, particularly relevant for large-scale, multi-institutional collaborations.[63] With the development of privacy-preserving technologies such as federated learning and secure multi-party computation (SMPC), researchers can collaboratively extract insights from sensitive datasets such as patient records or genomic data directly by processing encrypted or locally stored data and without transferring raw information beyond institutional or national boundaries.[64] Distributing databases across multiple nodes mitigates the risks of data loss, censorship, and unilateral control, while fostering transparency and trust in scientific outputs. This architecture shift reinforces the core tenets of open science, including collaboration, reproducibility, and fair access, by integrating resilience and inclusivity into the infrastructure that supports global research.

As science becomes increasingly data-intensive and globally interconnected, decentralization must be embedded into the fabric of scientific collaboration.[65] In an era shaped by AI, large-scale consortia, and cross-border data exchange, centralized infrastructures pose growing risks to the continuity, fairness, and transparency of research. Safeguarding open science in this context requires rethinking how data are stored, governed, and shared. Decentralized models offer a pathway to mitigate vulnerabilities associated with political, economic, or institutional disruptions, while promoting interoperability and trust across research communities.

Despite the promises of open science, structural inequities persist in who can contribute to and benefit from the global data infrastructure. Regions in various parts of the world, including parts of Eastern Europe, Central Asia, Sub-Saharan Africa, and the Caribbean, often face underrepresentation in federated networks. This is frequently due to limited national investments in digital infrastructure, institutional underdevelopment, or exclusion from established international consortia. For instance, while institutions in some transitioning economies have expressed interest in joining major international research infrastructures, a lack of sustained funding and policy alignment has often limited their full engagement. A truly fair system must invest in capacity-building initiatives that help underrepresented regions establish and manage their own interoperable nodes. These efforts should be supported by international funding mechanisms and reinforced by governance models that prioritize distributed authority and regional self-determination in data stewardship. Realizing this vision will require not only technical solutions but also policy alignment, sustained investment, and global coordination. To ensure that scientific knowledge remains a durable and accessible public good, decentralization must become a foundational principle in the design of future research infrastructures.

**Economic imperatives, principles, and challenges of building self-sustaining decentralized open-science databases:**

While DeSci offers a compelling vision for enhancing the resilience, transparency, and autonomy of biomedical data infrastructures, its implementation faces substantial practical and economic challenges, including scalability, cost-effectiveness, usability, governance, and legal interoperability. Current decentralized storage technologies face constraints when applied to large-scale biomedical data such as the SRA, where storage and retrieval demands may exceed the operational capacity and cost efficiency of current decentralized systems. In addition, the technical architecture of DeSci tools may diverge from the workflows and expectations of the life sciences community, where researchers typically rely on familiar web-based platforms and standardized APIs, in contrast to the specialized knowledge and infrastructure required by decentralized systems. Establishing effective governance and curation mechanisms in a decentralized environment, ensuring data quality and consistency in the absence of central authorities, also presents a significant challenge, as current decentralized autonomous organization (DAO)-based models are still in early stages of development. The legal uncertainty surrounding DAOs and decentralized data management further complicates implementation across diverse jurisdictions. Ultimately, the central question transcends technological feasibility and resilience, focusing instead on the economic sustainability of a decentralized system. Unlike centrally funded infrastructures or even federated models like ELIXIR, where costs can be distributed across national contributions, a truly DeSci approach necessitates a self-sustaining economic model where individual incentives for participating institutions/countries align with the collective goal and responsibility of maintaining the network.

Establishing a resilient, decentralized ecosystem for open-science data requires coordinated global action across multiple domains. First, a globally federated network merits prioritization, building on proven models such as ELIXIR to distribute responsibility and mitigate systemic risk. Rather than starting from scratch, this infrastructure could integrate existing strengths, drawing on the technical capabilities of NCBI, EMBL-EBI, and DDBJ, alongside the strategic coordination provided by the GBC. **(Figure 2)**. National and regional nodes may be supported through targeted capacity building and guided integration pathways to ensure coherent participation within a globally federated framework. Critical datasets may be redundantly mirrored across geographic regions through coordinated funding and replication agreements to ensure long-term preservation. An international governance body, potentially through the GBC, may establish standards, facilitate agreements, and oversee fair data access. While DeSci technologies hold promise for enhancing transparency and resilience (e.g., via immutable data logs or censorship-resistant archiving), their deployment should be guided by rigorous evaluation and integrated selectively where they add value. Finally, a unified global call to action is essential, mobilizing governments, funders, and institutions to invest in the foundational infrastructure required to protect and democratize scientific data for future generations.

The most pressing challenge in developing a truly decentralized open-science infrastructure is clearly defined and internationally trusted database governance, balancing diverse national interests while ensuring transparency and accountability. Such governance should promote and adopt rigorous, shared technical standards, spanning metadata, APIs, ontologies, and query protocols, coordinated by a neutral, expert authority, allowing deep interoperability. As distributed architectures increase vulnerability to cyber threats, necessitating harmonized security practices, robust access control, and synchronized data integrity mechanisms will go a long way. Long-term sustainability of such a system also hinges on reliable multi-national funding commitments and transparent financial governance. In the end, legal and ethical compliance must be assured through careful navigation of global data protection laws and ethical frameworks, especially when handling sensitive human data.

## Towards a resilient future for scientific data:

The decentralization of biomedical data infrastructure may mark a conceptual shift in the stewardship of knowledge, one that redefines how research outputs are preserved, governed, and shared. Beyond its technical implications, this transformation offers a framework for enhancing global data resilience, ensuring fair access, and securing the long-term sustainability of digital scientific assets. These benefits extend beyond individual researchers or institutions, shaping a more inclusive and durable foundation for the future of open science. However, technical, legal, political, and ethical complexities must be proactively addressed, including cybersecurity vulnerabilities, infrastructural disparities, and fragmented governance. A decentralized future for open-science databases cannot be realized through technological innovation alone; it demands coordinated solutions that integrate sustainable funding mechanisms, robust governance frameworks, and multilateral collaboration.

Among the three data-sharing architectures, each carries distinct liabilities: fully centralized repositories offer uniform standards but expose science to single-point failures, policy whims, and geographic inequities; federated networks disperse risk yet still struggle with uneven node resources, patchy interoperability, and fragmented governance; and decentralized (DeSci) models maximize resilience and transparency, but face steep technical costs and immature incentive structures (Figure 1).

A possible solution to address some or all of these issues lies in designing a Global Federated Architecture (GFA) that blends the strengths of all three approaches. As illustrated in Figure 2, this approach envisions autonomous national or institutional nodes retaining sovereignty and domain expertise, with a lightweight, consensus-based coordination layer enforcing common FAIR/CARE standards, metadata schemas, and security protocols. Cross-mirrored backups, reinforced by optional cryptographic audit trails, add DeSci-level tamper resistance. By coupling distributed governance with shared funding mechanisms and capacity-building for under-resourced regions, the GFA would decrease single-hub vulnerability, remove incompatibility gaps, and embed financial sustainability.

While a full migration of legacy databases to decentralized architectures isn't yet feasible, a hybrid architecture that selectively integrates DeSci principles into federated models may provide a balanced and future-ready solution (Figure 2). Such a layered approach would ensure the robustness and openness of scientific data, its ethical governance, long-term viability, and utility across an increasingly global and interdisciplinary research landscape.

Creating the GFA could begin with a coalition of organizations already promoting open science (e.g., ELIXIR, INSDC partners, GA4GH), major funders who support these resources, and representatives from regions often excluded from global infrastructure. This community would develop technical and policy standards and prepare the core infrastructure. Capacity-building and equity can be introduced at each phase, with training programs, shared services, and starter grants helping emerging nodes meet baseline technical and governance requirements. This system could evolve incrementally, with new nodes added and tested for compliance, standards, and interoperability. We believe it's time to start working towards the GFA. While the exact steps in building this global architecture can be refined, its overarching goals are clear: interoperable standards, distributed governance, sustainable funding, and an explicit commitment to equity are essential to ensure that scientific data remains a resilient, accessible, and truly worldwide public good. Safeguarding this shared scientific legacy demands coordinated action from researchers, institutions, funders, and policymakers to establish a resilient, transnational infrastructure that can endure political and economic uncertainty and preserve open knowledge for future generations.

**Acknowledgements**: Not applicable

**Consent for publication:** Not Applicable.

**Ethics approval and consent to participate:** Not applicable

**Availability of data and materials:** Not applicable

**Competing interests:** The authors declare no conflict of interest to disclose.

**Funding:** GSh acknowledges the funding from IIT Hyderabad, Science and Engineering Research Board, for supporting his research.

S.M., V.M., M.D. were supported by a grant of the Ministry of Research, Innovation and Digitization under Romania's National Recovery and Resilience Plan - Funded by EU – NextGenerationEU" program, project "Artificial intelligence-powered personalized health and genomics libraries for the analysis of long-term effects in COVID-19 patients (AI-PHGL-COVID)" number 760073/23.05.2023, code 285/30.11.2022, within Pillar III, Component C9, Investment 81.

Research reported in this publication was supported by the National Cancer Institute of the National Institutes of Health under Award Numbers U24CA248265 and U01AG066833.

The content is solely the responsibility of the authors and does not necessarily represent the official views of the National Institutes of Health or any other funding agencies.

**Authors' contributions:** Gaurav Sharma conceptualized the idea and wrote the initial draft. Gaurav Sharma and Serghei Mangul led the project and conceptualized its key components. Viorel Munteanu conceived, drafted, and revised the manuscript. Arunkumar Krishnan contributed to the early conceptual development of key components. All authors read, wrote, edited, and approved the final manuscript.

**Figure 1:** A side-by-side schematic that distills three progressively more distributed models of scientific data architecture. 1) Centralized systems: data, compute, and governance are consolidated within a single institutional locus, offering uniform standards but exposing science to "single-point-of-failure" risks and political or funding shocks. 2) Federated systems: nodes retain national or institutional autonomy but synchronize metadata, access policies, and core repositories through agreed-upon standards, distributing risk while preserving interoperability. 3) Decentralized (DeSci) systems: dispersed storage across many independent nodes, distributed file systems, and community-driven governance to create censorship-resistant and commonly shared self-sustaining data.

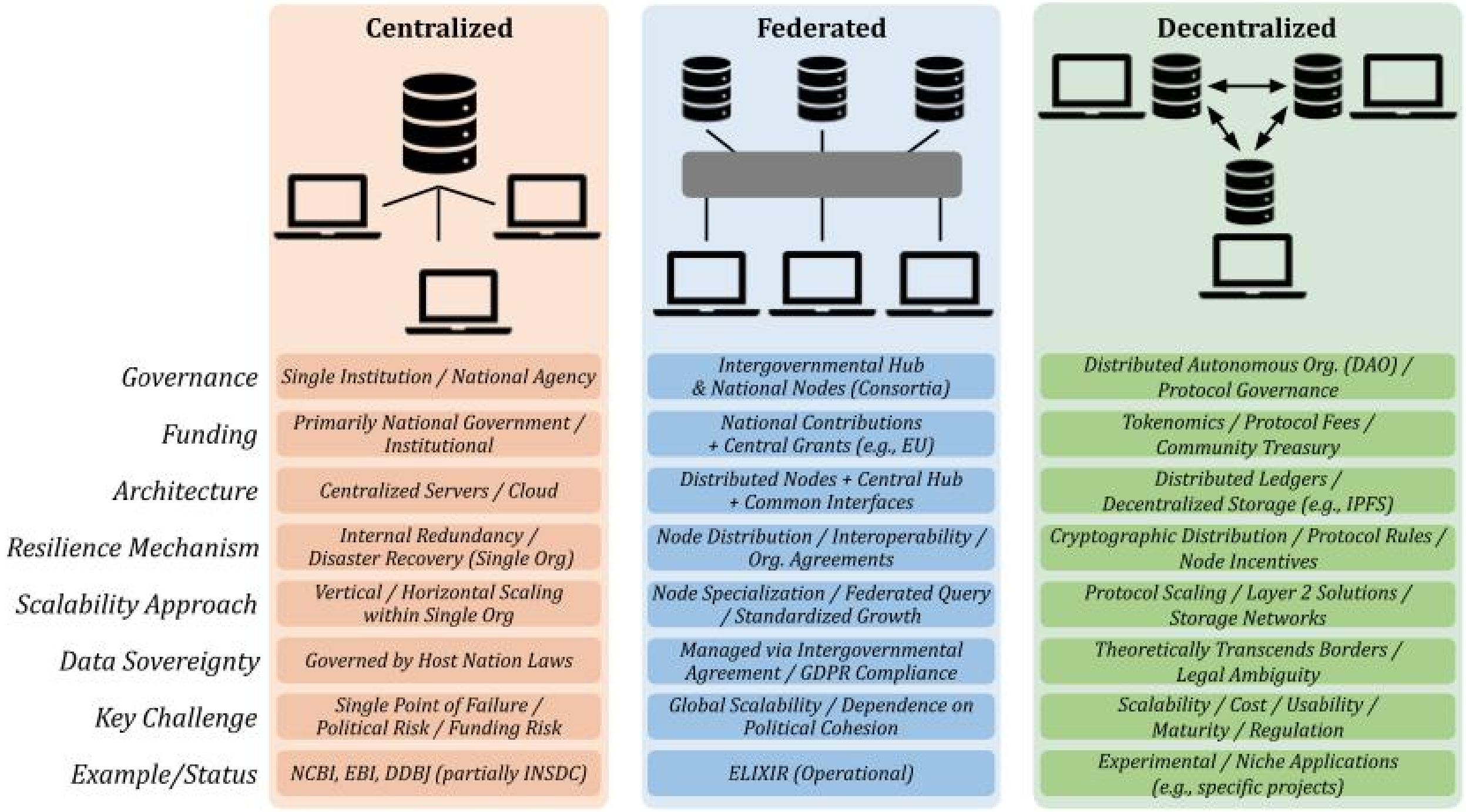

**Figure 2:** Conceptual flowchart of the Global Federated Architecture (GFA)—a proposed idea of an internationally distributed network of autonomous, standards-aligned data nodes coordinated by shared governance to maximize resilience, interoperability, and data sovereignty in open science. The model is particularly relevant to open science, healthcare, genomics, and other data-intensive disciplines where both data sovereignty and interoperability are essential.

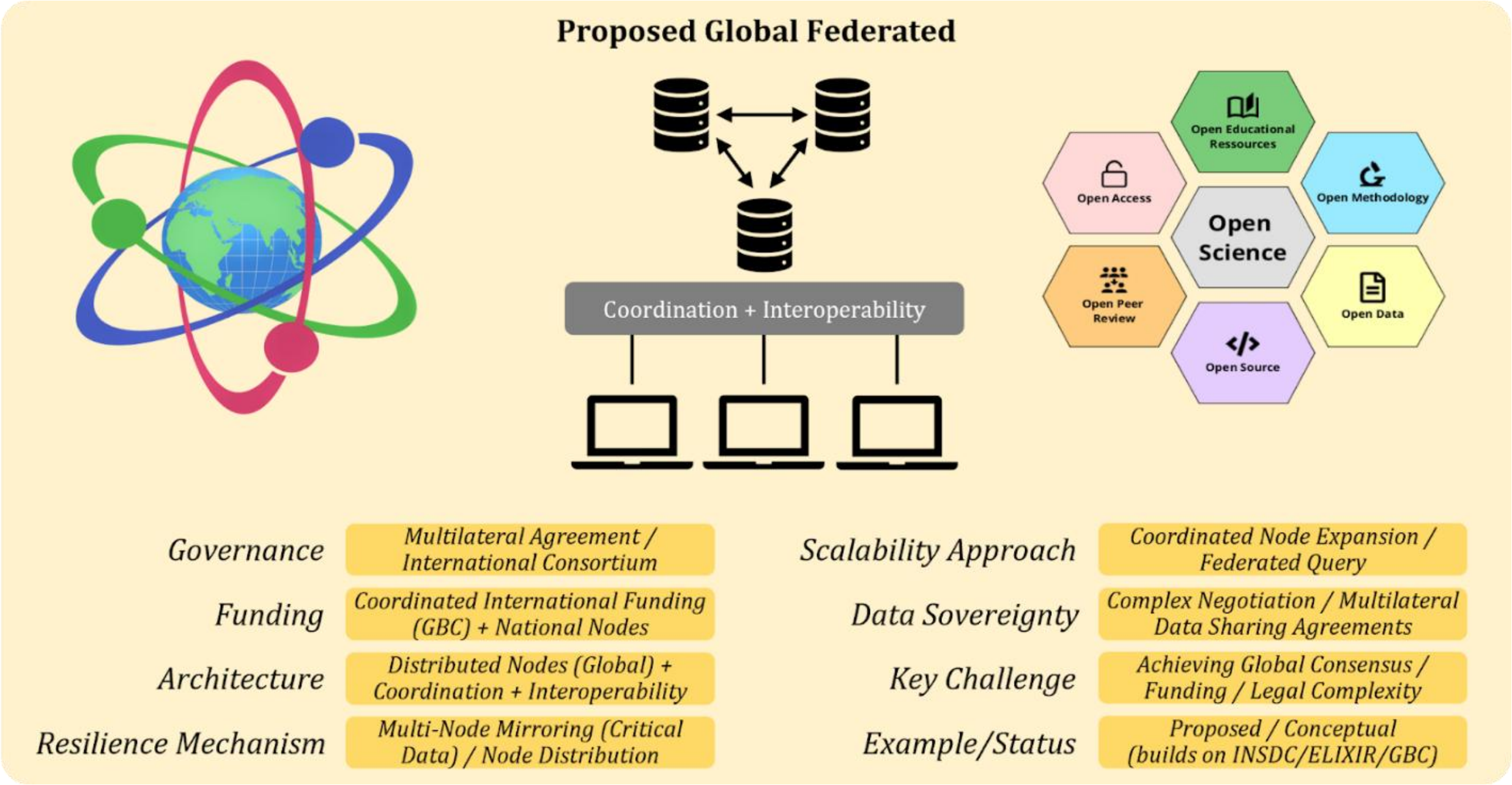